\documentclass[superscriptaddress,floatfix,twocolumn,showkeys,pra]{revtex4-2}
\usepackage{amsmath,amssymb,amsthm,mathrsfs,amsfonts,dsfont,mathtools}
\usepackage[normalem]{ulem}
\usepackage{bbm}
\usepackage{physics}
\usepackage{graphicx,wasysym}   
\usepackage{xcolor}
\usepackage{float}
\usepackage[ruled,vlined]{algorithm2e}
\usepackage{tcolorbox}
\tcbuselibrary{theorems}
\usepackage{nicefrac}
\usepackage{enumerate}
\usepackage{hyperref}
\usepackage[caption=false]{subfig}
\usepackage{tabularx}

\usepackage[capitalize]{cleveref}

\usepackage{xcolor}

\newcommand\Kappa{\mathrm{K}}

\newtcbtheorem[number within=section]{note}{Note}%
{colback=blue!5,colframe=blue!35!black,fonttitle=\bfseries}{th}

\usepackage{xifthen}
\newcommand{\PP}[1][]{
  \ifthenelse{\isempty{#1}}
    {\mathbbm{P}}
    {\mathbbm{P}\left[#1\right]}
}
\newcommand{\EE}[1][]{
  \ifthenelse{\isempty{#1}}
    {\mathbbm{E}}
    {\mathbbm{E}\left[#1\right]}
}

\newcolumntype{P}[1]{>{\centering\arraybackslash}p{#1}}

\begin{document}
\title{Preparing Ground and Excited States Using Adiabatic CoVaR}
\author{Wooseop Hwang}
\email{useop02@gmail.com}
\affiliation{Mathematical Institute, University of Oxford, Woodstock Road, Oxford OX2 6GG, United Kingdom}

\author{B\'alint Koczor}
\email{koczor@maths.ox.ac.uk}
\affiliation{Mathematical Institute, University of Oxford, Woodstock Road, Oxford OX2 6GG, United Kingdom}

\date{\today}

\begin{abstract}
CoVarince Root finding with classical shadows (CoVaR) was recently introduced as a new paradigm for training variational quantum circuits. Common approaches, such as variants of the Variational Quantum Eigensolver, aim to optimise a non-linear classical cost function and thus suffer from, e.g., poor local minima, high shot requirements and barren plateaus. In contrast, CoVaR fully exploits powerful classical shadows and finds joint roots of a very large number of covariances using only a logarithmic number of shots and linearly scaling classical HPC compute resources. As a result, CoVaR has been demonstrated to be particularly robust against local traps, however, its main limitation has been that it requires a sufficiently good initial state. We address this limitation by introducing an adiabatic morphing of the target Hamiltonian and demonstrate in a broad range of application examples that CoVaR can successfully prepare eigenstates of the target Hamiltonian when no initial warm start is known. CoVaR succeeds even when Hamiltonian energy gaps are very small -- this is in stark contrast to adiabatic evolution and phase estimation algorithms where circuit depths scale inversely with the Hamiltonian energy gaps. On the other hand, when the energy gaps are relatively small then adiabatic CoVaR may converge to higher excited states as opposed to a targeted specific low-lying state. Nevertheless, we exploit this feature of adiabatic CoVaR and demonstrate that it can be used to map out the low lying spectrum of a Hamiltonian which can be useful in practical applications, such as estimating thermal properties or in high-energy physics.
\end{abstract}

\maketitle

\section{Introduction\label{section:introduction}}

One of the most natural applications of quantum computers is
the simulation of quantum systems which, for certain tasks, promises exponential speedups over
classical computers. However, it is anticipated that early generations of quantum
computers (NISQ and early fault tolerant) will be limited in the numbers of qubits
and in the circuit depths that can be implemented~\cite{Preskill_2018}.
This motivated the field to consider hybrid algorithms whereby a quantum computer
is used to run shallow quantum circuits whose outputs are post processed 
in classical computers.
Such variational quantum algorithms~~\cite{vqeoriginal, Endo_2021, Cerezo_2021} 
have received significant attention and have been extended to a broad range of
application areas such as simulating time evolution~\cite{PhysRevX.7.021050}
or finding ground and excited states~\cite{Higgott_2019,oled,imaginary,PhysRevA.106.062416}.

On the other hand, variational quantum algorithms suffer from significant
performance limitations. First, estimating the cost function and its gradient to high
accuracy may require a prohibitively large number of shots~\cite{Endo_2021, Cerezo_2021, TILLY20221}.
Second, the optimisiation landscape is particularly complex and training is
generally NP-hard due the presence of a large number of local
optima~\cite{PhysRevLett.127.120502, Anschuetz_2022}.
Furthermore, if no good initial parameters are known then 
the optimisation may require exponential efforts to estimate gradients
due to barren plateaus~\cite{barrenplateau_qcnn}.

Recently, ref.~\cite{Boyd_2022} introduced the CoVariance Root finding (CoVaR) approach,
which fundamentally redefines the problem of training circuit parameters
by posing the problem as finding joint roots of a large number of covariance functions
rather than the minimisation of a classical cost function.
The approach is powerful as it can fully exploit classical shadows~\cite{Huang_2020} 
to estimate a large number of covariance functions using only a logarithmic number of
shots and using linearly scaling, but substantial HPC classical compute resources. The approach has the significant benefit that it can avoid local traps
and is robust against imperfections, such as gate noise~\cite{Boyd_2022}.

However, the main limitation of the CoVaR approach is its sensitivity to good initial parameters, i.e.,
ref.~\cite{Boyd_2022} numerically observed that CoVaR---similarly to phase estimation
algorithms---randomly converges to one of the nearby eigenstates
with a probability approximately proportional to the overlap in the initial state. While CoVaR has been a promising approach for training circuits when good initial parameters are available, in 
the present work we significantly extend its capabilities to the practically more important case
when no good initial states are known. As such, we demonstrate in a broad range
of application examples that CoVaR
can successfully prepare eigenstates of systems by initialising in an exact eigenstate of an initially
solvable Hamiltonian which is then gradually morphed into the desired problem Hamiltonian.
The only downside, however, is that in general one does not have full control
as to which eigeinstate the
approach converges to --  albeit we can guarantee convergence to specific eigenstates when 
the ansatz is sufficiently deep and the relevant energy gaps in the Hamiltonian are sufficiently large.

The present approach, called adiabatic CoVaR, uses a time-dependent Hamiltonian $\mathcal{H}(t)$
and builds on the concept of adiabatic computing, such that
one initialises in an eigenstate of a trivial Hamiltonian at $\mathcal{H}(t=0)$
and then gradually interpolates in steps of $\Delta t$ until the final problem Hamiltonian
is approached at $\mathcal{H}(t=1)$. At each step CoVaR is used to prepare an
instantaneous eigenstate of the formally time-dependent Hamiltonian and given a sufficiently small $\Delta t$, 
the initial state supplied to CoVaR is always guaranteed to have a
high overlap with a targeted instantaneous eigenstate.

We numerically demonstrate the effectiveness of this 
approach in a broad range of application examples, including finding ground and excited states
of spin problems, finding solutions to combinatorial optimisation problems and solving 
Hamiltonian eigenstates in high-energy physics. While VQE could, in principle, be used as part of
this approach to find instantaneous eigenstates, in all present numerical simulations in the 
present work VQE got trapped in local optima.
In contrast, in all numerical demonstrations
in the present work we found that CoVaR successfully converged to an eigenstate of the
problem Hamiltonian providing strong evidence for its robustness against local traps. 
However, the main limitation of the approach is that small energy gaps and shallow ansatz
circuit depths force CoVaR to converge to higher excited states even when it was initialised
in a low-lying state of the initial model. Nevertheless, this may
be a valuable feature of the present approach as adiabatic CoVaR allows to map out the low lying spectrum
of a Hamiltonian -- as we demonstrate this can benefit high-energy physics applications.

Crucially, while the complexity of conventional adiabatic evolution and phase estimation
protocols depend inversely on Hamiltonian energy gaps,
we numerically observe that adiabatic CoVaR has a fundamentally improved
robustness against small energy gaps and empirically observe that the step size $\Delta t$
has a logarithmic dependence on the smallest Hamiltonian energy gap within the
finite regime investigated. Furthermore, the parameter $\Delta t$ does not
affect the precision of the final eigenstate preparation, nor the circuit depths required,
and we demonstrate
that CoVaR can prepare eigenstates to relatively high precision even when using
relatively large $\Delta t$ -- this is in stark contrast to phase estimation and adiabatic evolution
algorithms whereby the circuit depth needs to scale inversely with the energy gaps 
otherwise the approaches cannot output an eigenstate. Of course, as we demonstrate, adiabatic
CoVaR potentially outputs higher excited states when $\Delta t$ is too large.
   
This manuscript is organised as follows.
In the rest of this introduction we briefly review basics of
preparing eigenstates using variational circuits and using adiabatic
evolution. Then in \cref{sec:adiabatic_CoVar} we present details of our
adiabatic CoVaR approach and then demonstrate its the effectiveness
in a range of practically important application examples in \cref{sec:appl}.
Finally, we demonstrate the robustness of the present approach against imperfections
and against different hyper parameter choices and then conclude.


\subsection{Preliminaries: Training Variational circuits}

\noindent \textbf{Variational quantum circuits} are constructed as a sequence of $L$
gate operations that are often chosen to be native, parametrised
gate operations of the hardware platform~\cite{Cerezo_2021} as
$U(\boldsymbol{\theta}) =  U_{L}(\theta_{L}) \dots U_{2}(\theta_{1})U_{1}(\theta_{1})$.
These circuits are then applied to an initial state, such as $\ket{0 \dots 0}$, to prepare a parametrised family of variational states as
\begin{align}
    \ket{\psi(\boldsymbol{\theta})} =  U(\boldsymbol{\theta})\ket{0 \dots 0}.
\end{align}
The particular construction of an 'ansatz' circuit $U(\boldsymbol{\theta})$ 
is often motivated by physical considerations, e.g., in the case of the Hamiltonian Variational Ansatz the individual gates in the ansatz circuit correspond to Pauli terms in the problem Hamiltonian~\cite{Cerezo2021}.

A common task in quantum simulation is to find the ground state
of a problem Hamiltonian  
$\mathcal{H} = \sum_{a} w_a P_a$
that is specified as a linear combination of Pauli strings
 $P_a \in \{ \openone, X, Y, Z \}^{ \otimes N }$. 
In Variational Quantum Eigensolvers (VQE), the expected value of this Hamiltonian
$C(\boldsymbol{\theta}) = \bra{\psi(\boldsymbol{\theta})} \mathcal{H} \ket{\psi(\boldsymbol{\theta})}$
is minimised as a cost function according to the variational principle
such that the global minimum over the parameters $\boldsymbol{\theta}$
approximates the ground state~\cite{PhysRevA.106.062416, vqeoriginal,TILLY20221}.

While the approach may be effective at moderate system sizes,
as one scales up, the classical optimization procedure may require exponential training efforts
as in general the VQE optimization problem is NP-hard~\cite{PhysRevLett.127.120502}.
In particular, if no good initial guess of the optimal parameters is known
then training may become prohibitively expensive due to barren plateaus~\cite{barrenplateau_qcnn,larocca2024review}. 
However, even if good initial states are know, it has been shown that the optimization
landscape of VQE contains exponentially many local traps~\cite{Anschuetz_2022}.
This motivates alternative techniques for training variational circuits,
and as we demonstrate in the present work,  adiabatic CoVaR mitigates both issues: the parameters are guided by an adiabatic
evolution and thus the system is always ``warm started'' while getting trapped in local optima is 
mitigated through the use of CoVaR.

\noindent \textbf{Covariance Root Finding (CoVaR)} 
was introduced in~\cite{Boyd_2022};
The approach finds eigenstates through a root finding problem over a very large
number of covariance surfaces rather than through the minimisation of a single energy surface as in VQE.
In particular, given any two Hermitian operators $A$ and $B$ one can define the covariance between
them that depends on a specific input quantum state $\ket{\psi}$ 
(which is assumed to be pure for ease of notation~\cite{Boyd_2022}) as
\begin{align}
    \langle A , B \rangle_{\psi} \coloneq \bra{\psi} AB \ket{\psi} - \bra{\psi}A\ket{\psi}\bra{\psi}B\ket{\psi}.
\end{align}
We choose these operators from a predefined operator pool
$\mathcal{P} \coloneq \{O_k\}^{r_p}_{k=1}$
that in the present work we assume to be local Pauli strings
$O_k \in \{\openone, X, Y, Z\}^{\otimes N}$, but any other orthonormal 
operator basis can also be used~\cite{Boyd_2022}. 

The approach builds on the main observation that 
given our problem Hamiltonian
$\mathcal{H} = \sum_{a}^{r} h_a  \mathcal{H}_{a}$, such that $\mathcal{H}_{a} \in \mathcal{P}$,
then a quantum state $\ket{\psi}$ 
is an eigenstate of $\mathcal{H}$ 
if the following conditions are met as
\begin{align*}
    \text{Sufficient condition:}  \quad  &\langle \mathcal{H}_a, \mathcal{H} \rangle_{\psi} = 0 \quad \forall \mathcal{H}_a \in \mathcal{P}, \\ 
    \text{Necessary condition:}  \quad &\langle O_k, \mathcal{H} \rangle_{\psi} = 0 \quad for \quad O_k \in \mathcal{P}.
\end{align*}
Then, given a parametrised variational quantum state
$\ket{\psi (\boldsymbol{\theta})} = U(\boldsymbol{\theta}) \ket{0}^{\otimes N}$,
the parameterised covariance is defined as~\cite{Boyd_2022}:
\begin{align}
	f_{k}(\boldsymbol{\theta}) \coloneq \langle O_k, \mathcal{H} \rangle_{\psi (\boldsymbol{\theta})}.
\end{align}
Indeed, similarly to VQE cost functions~\cite{Koczor_2022}, the
covariance function $f_{k}(\boldsymbol{\theta})$ is an infinitely
differentiable function of the circuit parameters
$\boldsymbol{\theta}$ for any Hermitian operator
$O_k$ and Hamiltonian $\mathcal{H}$.

This converts the problem of finding eigenstates into the problem of finding
roots of covariances over the parameters, i.e, one searches for parameter values $\boldsymbol{\theta}$ 
such that covariances in the state $\ket{\psi(\boldsymbol{\theta})}$
are simultaneously all zero for all operators in the operator pool.
There exist numerous methods for finding 
simultaneous roots of such vector-valued functions
as variants of Newton's  method~\cite{optimization};
One iterates the parameter values according to the update rule as

\begin{align}
    \boldsymbol{\theta}_{t+1} = \boldsymbol{\theta}_t - \boldsymbol{J}^{-1}\boldsymbol{f}.
\end{align}
Here both the vector of covariances $\boldsymbol{f}$
and the Jacobian $\boldsymbol{J}$, which is a matrix of all partial derivatives of
$\boldsymbol{f}$, are estimated using a quantum computer and they are defined as
\begin{align}
	\label{eq:f}
	\boldsymbol{f}_k(\boldsymbol{\theta}) = \bra{O_k} \ket{\mathcal{H}}_{\psi(\boldsymbol{\theta})},
	\quad \quad [\boldsymbol{J}]_{kl} = \partial_{k} \boldsymbol{f}_l(\boldsymbol{\theta}).
\end{align}

The significant advantage of CoVaR is that we can use a very large operator pool
as we can simultaneously estimate a large number of covariances using classical shadows.
For ease of notation, we focus on operator pools of local Pauli strings and corresponding efficient Pauli shadows, however, as we detail
below a range of more advanced classical shadow techniques are immediately applicable.
Using these local Pauli shadows, we first prepare the variational state using our ansatz circuit
$\ket{\psi(\boldsymbol{\theta})}$
and then measure each qubit randomly in either the $X$, $Y$, or $Z$ bases
(by applying random single-qubit rotations)
which results in an $N$-bit classical measurement outcome as the bitstring $\ket{b_i} \in \{ 0, 1\}^N$.
The collection of measurement bases and outcome bitsrings forms a classical dataset which is the
classical shadow of the quantum state $\ket{\psi(\boldsymbol{\theta})}$~\cite{Huang_2020}.
All covariances can then be estimated efficiently by post-processing this classical shadow dataset.

A remarkable and unique feature of CoVaR is that it is robust against local traps
in the energy surface: ref~\cite{Boyd_2022} demonstrated VQE optimisations
that get trapped in local optima; initialising CoVaR in such a state, the approach
can ``jump out'' of these traps.

In fact, CoVaR 
can be compared with phase estimation algorithms, as the success rate of converging
to the desired eigenstate was found proportional to the initial state overlap in numerical simulations.
On the other hand, the main limitation of CoVaR was found that it fails to make progress when
initialised away from an eigenstate, i.e., when no good initial parameters are known.
In the present work we demonstrate adiabatic CoVaR in which the CoVaR optimiser is 
used iteratively and is always inputted an initial state that is
close to the desired target state.


\subsection{Adiabatic evolution}

In adiabatic quantum computation one prepares the ground state of a trivial Hamiltonian
and slowly morphs this Hamiltonian into the true problem Hamiltonian such that
the system remains in the instantaneous ground state of the time-dependent Hamiltonian~\cite{adiabatic,farhi2000quantum}.

For example, one assumes a Hamiltonian $H_a$ 
is available whose ground state is known 
and the aim is to find the ground state of the Hamiltonian $\mathcal{H}_b$.
The adiabatic approach is then typically applied to the linear combination as
\begin{align}
	\label{eq:adiabatic_Hamil}
	\mathcal{H}(s) = (1-s)\mathcal{H}_0 + s \mathcal{H}_1.
\end{align}
In general, given a one-parameter family of Hamiltonians
$\mathcal{H}(s)$ for $0 \leq s \leq 1$ 
we can identify its instantaneous ground-state via the eigenvalue equation
\begin{align}
    \mathcal{H}(s) \ket{\psi_0(s)} = E_0 (s) \ket{\psi_0(s)}.
\end{align}
Then, choosing the initial state $\ket{\psi_0(0)}$ as the ground state of the
initial trivial Hamiltonian $H(0)$,
the adiabatic theorem guarantees
that for a sufficiently long time evolution, the final state will be
arbitrarily close to $\ket{\psi_0(s=1)}$
as long as the system is gapped
as $E_{1}(s) - E_{0}(s) > 0$ at all times $0 \leq s \leq 1$.

Setting $s=t/T$, where $t$ is a physical time variable, the total evolution time $T$ needs to be chosen according to
the condition as
\begin{align}
	\label{adiabaticTime}
	T \gg \frac{\epsilon}{g_{min}^{2}},
\end{align}
which guarantees that the time-evolved state stays close
to the instantaneous ground state $\ket{\psi_{0} (s)}$ throughout the evolution. 
Here the  minimum energy gap is defined as
\begin{align}
    g_{min} = \min_{0 \leq s \leq 1} [E_1 (s) - E_{0}(s)],
\end{align}
while $\epsilon$ depends on the transition amplitudes as

\begin{align}
\label{adiabaticE}
    \epsilon = \max_{0 \leq s \leq 1} = | \bra{ \psi_{1}(s)}
     \frac{\mathrm{d} \mathcal{H}}{\mathrm{d} s} \ket{\psi_0(s)} |.
\end{align}

The time evolution under the parametrised Hamiltonian can be simulated
using a range of techniques, including trotterisation, 
variational quantum simulation~\cite{Yuan_2019}, quantum walk-based methods ~\cite{Berry_2015},
Taylor series expansions~\cite{7354428}, and qubitization~\cite{Low_2019}.
While adiabatic computing is indeed a general and powerful approach, 
its implementation may be challenging due to deep quantum circuits required
for simulating sufficiently long time evolutions. While our approach is indeed inspired by 
adiabatic evolution, we do not use time evolution but rather use CoVaR to train a parametrised short-depth
circuit to follow instantaneous eigenstates of the problem Hamiltonian.

\section{Adiabatic CoVaR}\label{sec:adiabatic_CoVar}

We now detail our approach which combines an adiabatic evolution, i.e., gradual 
morphing of a Hamiltonian, and CoVaR which is used to train a shallow-depth parametrised circuit
in order to closely follow instantaneous eigenstates of the adiabatic Hamiltonian.
The significant advantage of this approach is that no time evolution is used but rather
parameters of a shallow ansatz circuit are slowly varied using CoVaR. While CoVaR has been
observed in numerical simulations to rapidly and robustly converge to eigenstates,
its main limitation has been that it needs an initial 
state that has sufficient overlap with the target eigenstate. The present
adiabatic morphing guarantees that CoVaR is always supplied a initial state that has high overlap with 
an instantaneous eigenstate.

\subsection{Hamiltonian morphing \label{sec:ham_morphing}}
In the present work we explore two different Hamiltonian morphing techniques
both of which interpolate between an initial and the final Hamiltonians by gradually 
increasing a time parameter $t\in \{0, \Delta t, 2\Delta t, \dots, 1\}$
in discrete time steps of $ \Delta t$ that we will refer to as the adiabatic time-step.

First, we consider the usual adiabatic evolution Hamiltonian from \cref{eq:adiabatic_Hamil}
by setting $T \equiv 1$ as 
\begin{align}
	\label{eq:timeHamil2}
	\mathcal{H}(t) = (1-t)\mathcal{H}_0 + t\mathcal{H},
\end{align}
where $\mathcal{H}_0$ is an initial Hamiltonian of arbitrary choice that is easy to solve analytically
and $\mathcal{H}$ is the problem Hamiltonian of interest.
We will refer to this as the 'mixing approach'.

Second, we consider a special case that can be applied when the Hamiltonian 
decomposes into a---potentially dominant---contribution $\mathcal{H}_0$ which can be solved analytically
and into additional terms as $\mathcal{H} = \mathcal{H}_0 + \mathcal{H}_1$~\cite{Wecker_2015}.
We then consider the adibatic Hamiltonian as
\begin{align}
\label{eq:timeHamil1}
    \mathcal{H}(t) = \mathcal{H}_0 + t\mathcal{H}_1.
\end{align}
This variant is particularly relevant to physical systems where non-trivial perturbative interactions  $\mathcal{H}_1$
are added to an otherwise solvable system $\mathcal{H}_0$, and thus we will refer to
this as the 'perturbative approach'.

For example, we will later consider the spin ring Hamiltonian  in \cref{eq:spinring}
for which we can define $\mathcal{H}_0 = \sum_{i=1}^{N} c_i Z_i$
and $\mathcal{H}_1 =  J \sum_{i=1}^{N} \boldsymbol{\sigma_{i}} \cdot \boldsymbol{\sigma_{i+1}}$,
where $J$ is a coupling constant, $\sigma_{i}$ are vectors of single-qubit Pauli matrices
acting on the $i$-th qubit, and $-1 \leq c_i \leq 1$ are uniformly randomly generated onsite 
interactions.
This naturally leads to the parametrisation of the Hamiltonian as
\begin{equation}
     \mathcal{H}(t) = \sum_{i=1}^{N} c_i Z_i + t J \sum_{i=1}^{N} \boldsymbol{\sigma_{i}} \cdot \boldsymbol{\sigma_{i+1}}.
    \label{eq:spin_time_Hamil}
\end{equation}

The above decomposition is convenient as one can analytically solve for the eigenstates of the 
initial trivial model $\mathcal{H}_0$.
However, for systems in which such separation is not possible, the mixing approach should be employed.
In cases when $J$ is small, the perturbative approach \cref{eq:timeHamil1} 
may be preferable over the mixing approach \cref{eq:timeHamil2} as the initial 
state is merely perturbed by the additional non-trivial term $\mathcal{H}_0$, 
whereas the mixing approach transforms between Hamiltonians of entirely
different physical systems. 
An added benefit of the perturbative approach is that it prepares eigenstates
for the entire path $0 \leq J\leq 1$ not just for the final, target Hamiltonian with, e.g., $J=1$.

\begin{figure}[tb]
	\centering
	\includegraphics[width=0.47\textwidth]{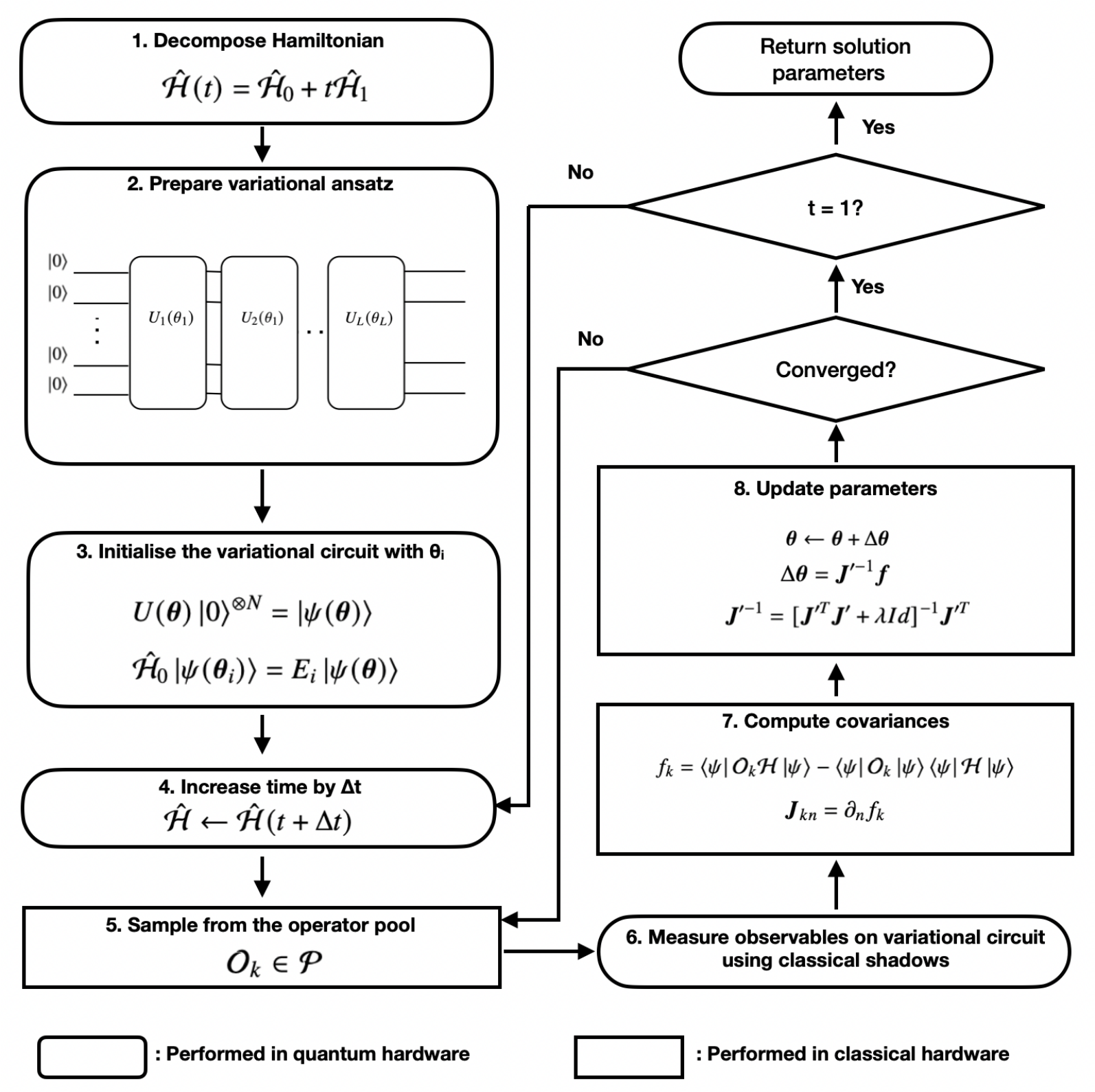}
	\caption{Flowchart of the Adiabatic CoVaR approach.
		\label{fig:adiabatic_flowchart}
	}
\end{figure}

\subsection{Adiabatic CoVaR}
\label{sec:adiabatic_CoVaR}

Our approach first initialises the variational circuit in an eigenstate of the initial Hamiltonian $\mathcal{H}(t=0)$.
The parameter value $t$ is then increased to $\Delta t$
and CoVaR is used to train the circuit parameters such that the variational circuit prepares an eigenstate of the
the Hamiltonian  $\mathcal{H}(\Delta t)$. This is then repeated iteratively until $t=1$ is reached.

More specifically, we use the variational ansatz
$\ket{\psi(\boldsymbol{\theta})} = U(\boldsymbol{\theta}) \ket{0}^{\otimes N}$,
in which the parameters $\boldsymbol{\theta} := \boldsymbol{\theta} (t, \kappa)$
depend on $t$ and $\kappa$. As illustrated in \cref{fig:adiabatic_flowchart}, $t$ is increased in steps
$t \in \{ 0, \Delta t, 2\Delta t, \dots, 1\}$
and for each fixed value of $t$, a CoVaR optimisation is run
whose iteration count is denoted by $\kappa$ and is capped at a maximum allowed CoVaR steps $\Kappa$.

The initial parameters $\boldsymbol{\theta} (t=0, \kappa=0)$ by construction
prepare an energy eigenstate of the initial Hamiltonian $\mathcal{H}_0$ as
\begin{align}
    \mathcal{H}_0 \ket{\psi(\boldsymbol{\theta}(0, 0))} = E_{i} \ket{\psi(\boldsymbol{\theta}(0, 0))}.
\end{align}

The parameters are then updated through the 	following update rules as
\begin{align}
    & \boldsymbol{\theta}(t + \Delta t, 0) =  \boldsymbol{\theta}(t , \Kappa) \\
    \label{eq:parameter_update}
    & \boldsymbol{\theta}(t, \kappa + 1) = \boldsymbol{\theta} (t, \kappa) - \boldsymbol{J}^{-1}\boldsymbol{f}(\boldsymbol{\theta}_i (t, \kappa)).
\end{align}
Here, $\boldsymbol{J}:= \boldsymbol{J}(\boldsymbol{\theta} (t, \kappa))$
is the Jacobian 
and $\boldsymbol{f}:= \boldsymbol{f}(\boldsymbol{\theta} (t, \kappa))$
is the covariance vector as defined in \cref{eq:f}.

A significant advantage of the approach is that it allows the direct preparation
of both the ground state and excited states using a short-depth variational ansatz -- albeit one does not have full
control over which eigenstate CoVaR converges to as we demonstrate below.
This is in contrast to VQD~\cite{Gocho2023,Higgott_2019}, which we summarise in \cref{app:vqd} and which requires the computation of
previous $n-1$ energy eigenstates for the preparation of $n$-th excited state.

\begin{figure*}[tb]
	\centering
 		\includegraphics[width=\textwidth]{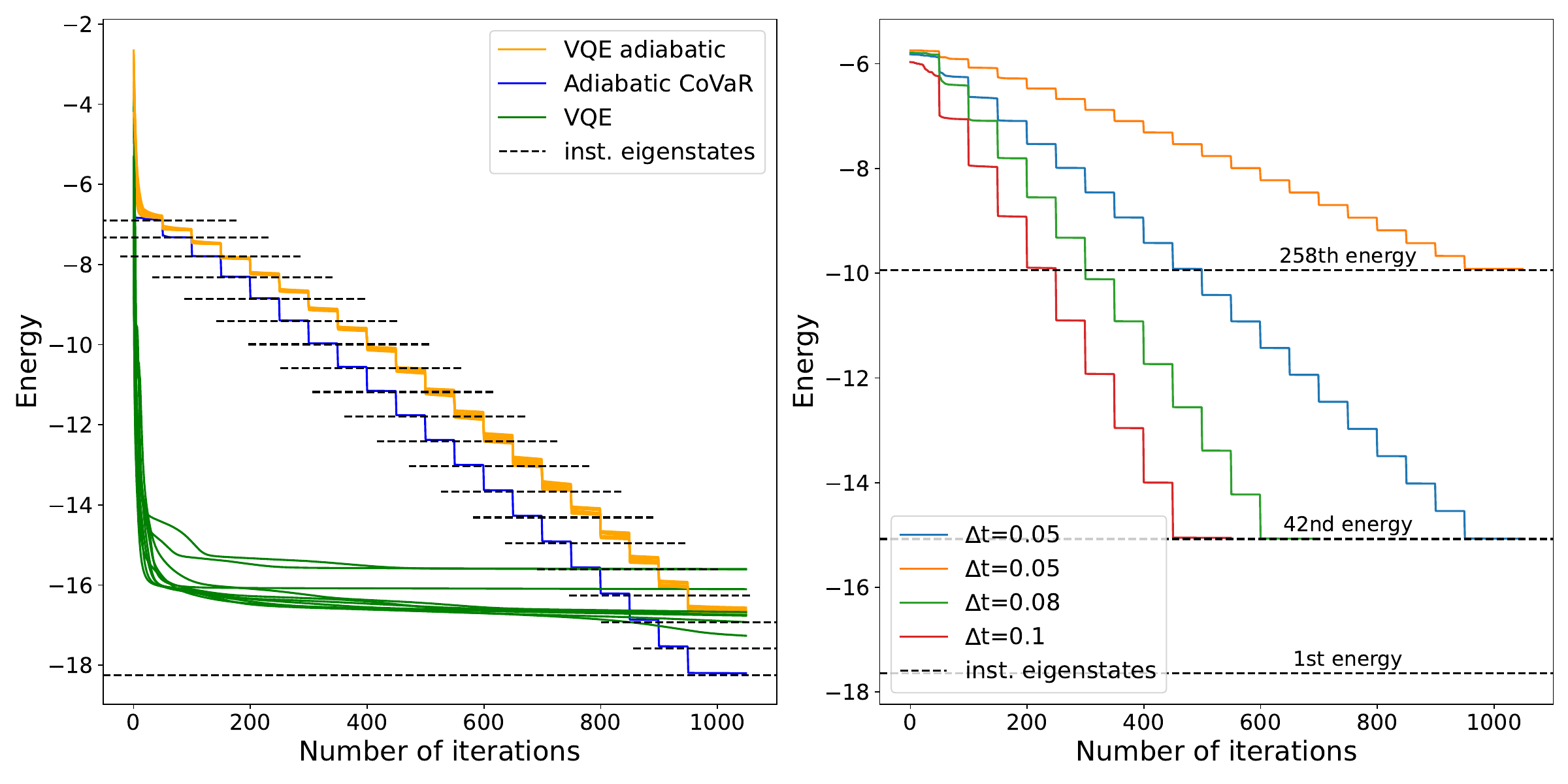} 
	\caption{
		 (left)
		 Preparing the ground state of a 10-qubit spin chain problem in \cref{eq:spinring} for an increasing coupling strength $0 \leq J\leq 1$ using
		 adiabatic CoVaR (blue).
		 Dashed black lines represent the instantaneous ground state
		 energies of the problem Hamiltonian (whose minimum energy gap was $g_{min}=0.742$). 
		 Blue approaches the dashed black lines (instantaneous ground-state energies) which confirms
		 that CoVaR could successfully prepare instantaneous ground states of the morphed Hamiltonian.
		 10 different runs of VQE applied to the same morphed Hamiltonian (orange) all get trapped
		 in local optima and thus cannot reach the target ground state.
		10 different runs of conventional VQE applied to the final target Hamiltonian $J=1$ (green) all get trapped.
		(right) preparing excited states of the same model for increasing time-steps $\Delta t$
		(red, green, orange, and blue) by initialising the CoVaR evolution in the first excited state of the initial model $\mathcal{H}(t=0)$.
		Given the spin model has a complex eigen-energy structure along the path $0 \leq t \leq 1$ with small energy gaps
		and energy crossovers, we illustrate in \cref{fig:level} that CoVaR gradually jumps to higher and higher excited states
		during the morphing evolution
		and finally ends up in the 42nd and 258th excited states of the final model (black dashed lines).
		While one has no control of which excited state CoVaR converges to, the approach indeed manages to prepare
		one of the excited states of the final model to relatively high precision $|\Delta E| \leq 0.022$.	
	\label{fig:CoVaR_vqe}}
\end{figure*}

The other significant advantage of adiabatic CoVaR is that a relatively large $\Delta t$
can be used compared to conventional adiabatic evolution.
As detailed in \cref{adiabaticTime}, the time complexity of adiabatic evolution is $O( g_{min}^{-2})$,
where $g_{min}$ is the minimum energy difference between the consecutive
eigenstates throughout the adiabatic evolution~\cite{adiabatic_scale}.

We will later demonstrate that indeed $\Delta t$ scales favourably with
the energy gap, and that $\Delta t$ does not significantly affect the precision of eigenstate preparation.
The present approach performs in total $K/(\Delta t)$ iterations of CoVaR
while ref~\cite{Boyd_2022} bounded the number of samples $N_s$, i.e., number of
circuit repetitions, required for each iteration: using classical shadows estimating
the Jacobian $\boldsymbol{J} \in \mathbb{C}^{N_c \times v}$ to an error
$\epsilon$ requires only a logarithmic overhead in the number $N_c$ of covariances
\begin{align}
    N_s = \mathcal{O} [ \nu \log(N_c) \epsilon^{-2}],
\end{align}
which allows us to estimate a very large number of covariances. The cost also depends proportionally
on the number $\nu$ of circuit parameters.

One of the other advantages we inherit from CoVaR is its robustness to gate noise, shot noise
and robustness to getting trapped in local optima as we are adopting a stochastic
Levenberg-Marquadt (LM) evolution~\cite{bergou2021stochastic, Liew2016},
whereby the operator pool $\mathcal{P}$, and thus the covariances 
are randomly chosen at each iteration as in~\cite{Boyd_2022}.

As we demonstrate later,
when the energy gap $g_{min}$ is too small, CoVaR may converge to a different eigenstate located close in parameter space. 
A possible approach to allow CoVaR to get back to the desired eigenstate
could be by adding small random fluctuations in the parameters $\boldsymbol{\theta}_i(t, 0)$
at every timestep $\Delta t$ -- or by occasionally performing VQE energy minimisation throughout the
evolution to force CoVaR to converge to low-lying eigenstates.


\section{Application examples}\label{sec:appl}

In this section, we present numerical simulations in a range of practically relevant application examples as 
(a) finding ground and excited states of a spin ring Hamiltonian,
(b) solving binary optimisation problems by finding the ground state of a Hamiltonian that encodes the solution to a max-cut problem
and
(c) finding excited states of the Schwinger model which is a toy model Hamiltonian used in high-energy physics applications.

\subsection{Eigenstate discovery of spin problems}

Spin models are important in a broad range of applications including condensed matter physics
and material science~\cite{Boyd_2022, Pagano_2020, RevModPhys.94.015004, Endo2021, Harrigan_2021}.
Furthermore, certain hard optimisations problems can be mapped to spin models~\cite{Lucas_2014}
and indeed certain spin models may be hard to simulate classically
for large numbers of qubits~\cite{PhysRevB.91.081103, Childs_2018}.
Conversely, finding ground states of certain spin problems for large $N$
may require exponentially increasing circuit depths~\cite{bosse2021probing}.

Here we consider the perturbative approach introduced in \cref{sec:ham_morphing}
to find eigenstates of the random field Heisenberg model as
\begin{align}
	\label{eq:spinring}
	\mathcal{H} =  \sum_{i=1}^{N} [  c_i Z_i +  J\boldsymbol{\sigma_{i}} \cdot \boldsymbol{\sigma_{i+1}}].
\end{align}
We fix the coupling strength $J=1.0$ to match the size of the onsite interactions
in which case the eigenstates are most complex and we choose a system size of $N=10$ qubits.
In the present perturbative morphing approach
we initialise the variational circuit to prepare the ground state of the
initial Hamiltonian $\mathcal{H}(0)$ as a computational basis state
and then gradually increase the coupling strength $J$ of the model as in \cref{eq:spin_time_Hamil}.

\cref{fig:CoVaR_vqe} (left, blue) shows the energy obtained with adiabatic CoVaR throughout the evolution
and confirms that, indeed, the present approach can closely follow the
instantaneous ground states
as the blue curves rapidly converge to the instantaneous
ground state energies of the problem Hamiltonian (black dashed lines). 
In fact, each time the blue line in \cref{fig:CoVaR_vqe} (left) 
converges to one of the dashed lines represents the successful preparation of the
ground state of the Hamiltonian at intermediate values of $J$ -- this nicely illustrates
that the present approach prepares an entire family of ground states for increasing values of $J$
as opposed to preparing only the ground state of the final Hamiltonian at $J=1$.

For comparison, in \cref{fig:CoVaR_vqe} (left, orange)  we apply a VQE evolution
to the same morphed Hamiltonian (effectively replacing the CoVaR subroutine with VQE).
In the present demonstration all 5 runs (initialised randomly around the same initial parameters) got
trapped in local optima significantly above the ground state energy -- indeed it is a known
limitation of VQE that it is vulnerable to getting trapped in local optima~\cite{localoptima}.
In contrast, the present adiabatic CoVaR approach clearly  converges to the true ground state
and achieves an energy difference $\Delta E = 0.055$ which is almost 2 orders of magnitude smaller.
We note that the energy difference of the CoVaR solution is 
limited by the depth of the ansatz circuit -- we use a coupling constant $J=1.0$
which requires a relatively expressive variational circuit to achieve a good representation
of the ground state and indeed CoVaR achieves significantly lower energy differences for the intermediate 
ground states at $J < 1$, for example,
$\Delta E$ is an order of magnitude smaller as $\Delta E \leq 10^{-3}$ for  $J \leq 0.4$.
Furthermore, we note that adiabatic CoVaR converges quickly to
the instantaneous ground states
at each time step as shown \cref{fig:CoVaR_vqe} (blue), and thus likely fewer total iterations could be used than 
$1050$ -- the relatively high iteration count was used to facilitate comparison with VQE.


\begin{figure*}[t]
	\centering
	\includegraphics[width=8.3cm]{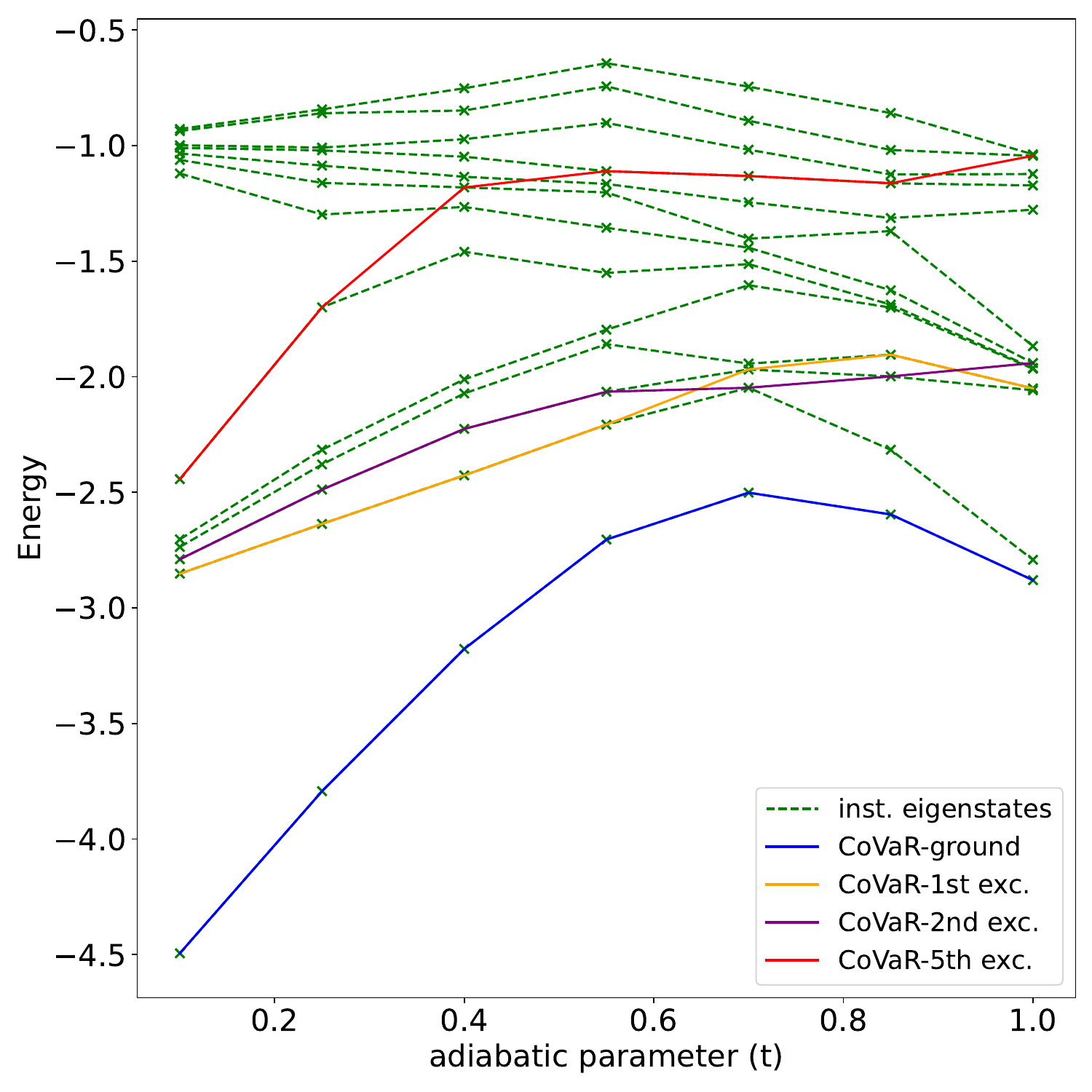}
	\includegraphics[width=8.3cm]{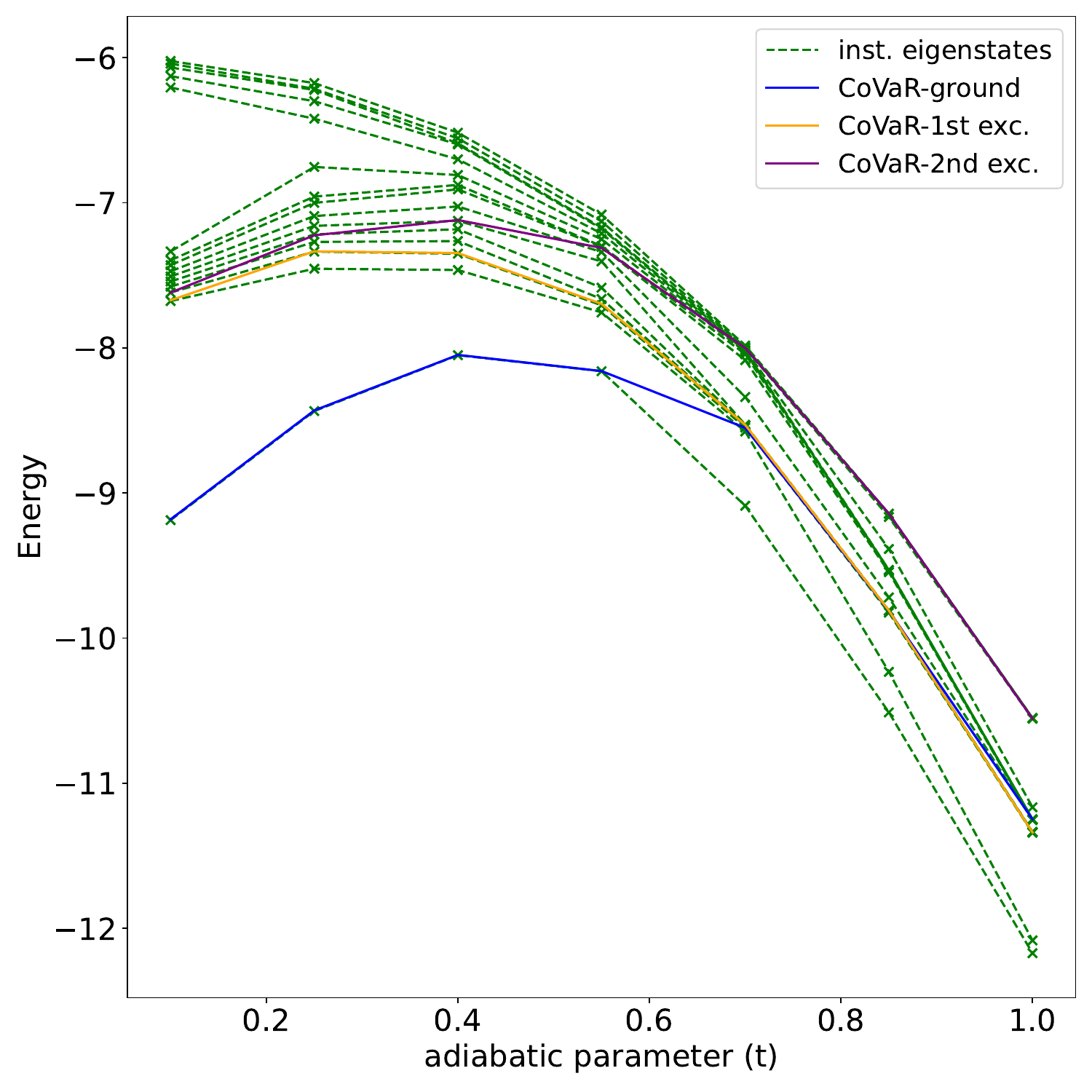}
	\caption{Energy Evolution in a 5-qubit (left) and 10-qubit (right) Lattice Schwinger Hamiltonian.
		Dashed lines are instantaneous eigenstates of the time-dependent Hamiltonian in \cref{eq:hepHamil_t} 
		using parameters  $J=1.0$, $m=0.1$, $\theta = 0$, and $w=0.1$.
		Solid lines indicate the evolution of the energy using adiabatic CoVaR 
		with a time step $\Delta t = 0.15$.
		Blue represents ground-state preparation by initialising in the ground state of the initial model. 
		Low-lying excited states of the final model were prepared by intialising in the first (orange) and second (purple) excited states of the initial model.
	}
	\label{fig:hep}
\end{figure*}


In \cref{fig:CoVaR_vqe} (left, green) we also compare to conventional
VQE and plot the evolution of the energy of 10 different rounds 
that were all initialised randomly in the vicinity of the ground state of the trivial model. 
While VQE can initially rapidly lower the energy, all runs get trapped in local
optima, far away from the true ground state as consistent with theoretical expectations~\cite{localoptima}.

Finally, we perform simulations for preparing excited states of  the present spin model, focusing only
on assessing the performance of CoVaR rather than comparing to VQE-based deflation methods.
\cref{fig:CoVaR_vqe} (right) shows adiabatic CoVaR evolutions for different time-steps 
$\Delta t \in \{ 0.01, 0.05, 0.08,0.1 \}$ when the evolution is intialised
in the first excited state of the initial trivial Hamiltonian $\mathcal{H}_0$.
If $\Delta t$ was very small and there were no energy-level crossovers then
the approach could prepare the first excited state of the final model. 
However, we will illustrate below in more detail that when $\Delta t$ is relatively large and small gaps are present,
CoVaR may converge to other energy eigenstates that are close in parameter space potentially resulting in excitations.
In  \cref{fig:level} we plot how adiabatic CoVaR gradually jumps to higher and higher excited states
and indeed \cref{fig:CoVaR_vqe} (right) demonstrates that following multiple consecutive jumps,
adiabatic CoVaR ends up preparing the 42nd and 258th excited states of the final model ($J=1$). 
Below \cref{fig:level} we also detail that both 
\cref{fig:CoVaR_vqe} (right, orange and blue)
use exactly the same hyperparameters, however, due its randomised nature,
adiabatic CoVaR ends up in a significantly higher excited state in \cref{fig:CoVaR_vqe} (right, orange).

Still, it appears to be a general feature of CoVaR that even though one cannot fully control which excited state 
to prepare, at least CoVaR prepares a very good approximation to one of the eigenstates as $\Delta E = 0.007$ (blue),
$\Delta E = 0.022$ (orange), $\Delta E = 0.007$ (green), and $\Delta E = 0.021 $ (red).
While we only feature excited states of the final model ($J=1$) in \cref{fig:CoVaR_vqe} (right, dashed lines),
CoVaR indeed prepares excited states of the parametric model along the entire path $0 < J \leq 1$. 
We can also infer this from \cref{fig:level}, which confirms that the present approach prepares
a range of low and intermediate excited states
for the entire family of Hamiltonians for increasing $J$.
A potential application as estimating thermal properties could be enabled by running CoVaR on multiple excited initial states
in order to map out the entire low-energy eigenspace of a Hamiltonian.

Finally, we perform further numerical analysis in \cref{sec:step&layer} and \cref{sec:step&iter},
whereby we systematically explore how different hyperparameters affect the performance of the present approach.
These results indicate that the number of variational layers can significantly affect the number of iterations required to reach a desired precision.
The reason is that choosing a low number of layers results in a
subspace spanned by the variational ansatz that is not sufficiently expressive to contain
the relevant eigenstates of all instantaneous Hamiltonians. 
In \cref{sec:ediff_evol} we also track the evolution of the imprecision $\Delta E$ 
throughout the adiabatic path. While we find that in the particular system simulated, ground-state 
preparation was rather stable, the imprecision monotonically increased throughout the evolution path
suggesting an increased complexity of preparing excited states.

\subsection{High-energy physics applications}

Here we consider a simple toy model in High-Energy Physics following ref.~\cite{chakraborty2022classicallyemulateddigitalquantum}
whereby the lattice Schwinger model was mapped to a spin Hamiltonian by enforcing Gauss' law 
and following a Jordan-Wigner transformation as
$\mathcal{H}_{sch} = \mathcal{H}_{ZZ} + \mathcal{H}_{\pm} + \mathcal{H}_{Z}$
where the individual Hamiltonian terms are defined as
\begin{align*}
	\mathcal{H}_{ZZ} &= \frac{J}{2} \sum_{n=2}^{N-1} \sum_{1\leq k \leq l \leq n} Z_k Z_l, \\
	\mathcal{H}_{\pm} &= \frac{J}{2} \sum_{n=1}^{N-1} [w-(-1)^n\frac{m}{2} \sin (\theta)][X_nX_{n+1} + Y_nY_{n+1}], \\
	\mathcal{H}_{Z} &= \frac{m \cos\theta}{2} \sum_{n=1}^{N} (-1)^nZ_n-\frac{J}{2}\sum_{n=1}^{N-1} (n\mod2)\sum_{l=1}^{n}Z_l.
\end{align*}
Ref.~\cite{chakraborty2022classicallyemulateddigitalquantum} then used this Hamiltonian 
to prepare massless and massive eigenstates of the lattice Schwinger model
using digital quantum algorithms. 
Here we consider an adiabatic mixing approach via the following time dependent Hamiltonian as
\begin{align}
\label{eq:hepHamil_t}
    \mathcal{H} (t) = (1-t)\sum_{i}^{N}X_i + t\mathcal{H}_{sch}.
\end{align}

In \cref{fig:hep} we perform adiabatic CoVaR on a 5-qubit (left) and on a 10-qubit (right) instance
of this Hamiltonian.
First, we initialise our approach in the ground state of the trivial mixing Hamiltonian
and plot the evolution of the energies obtained via CoVaR in \cref{fig:hep} (blue).
As the model has a significant ground-state energy gap for a the smaller system (left), our approach
manages to very closely follow the true ground state, however, for the larger system size (right) CoVaR eventually
converges to the first excited state.

\begin{figure*}[t]
	\centering\includegraphics[width=\textwidth]{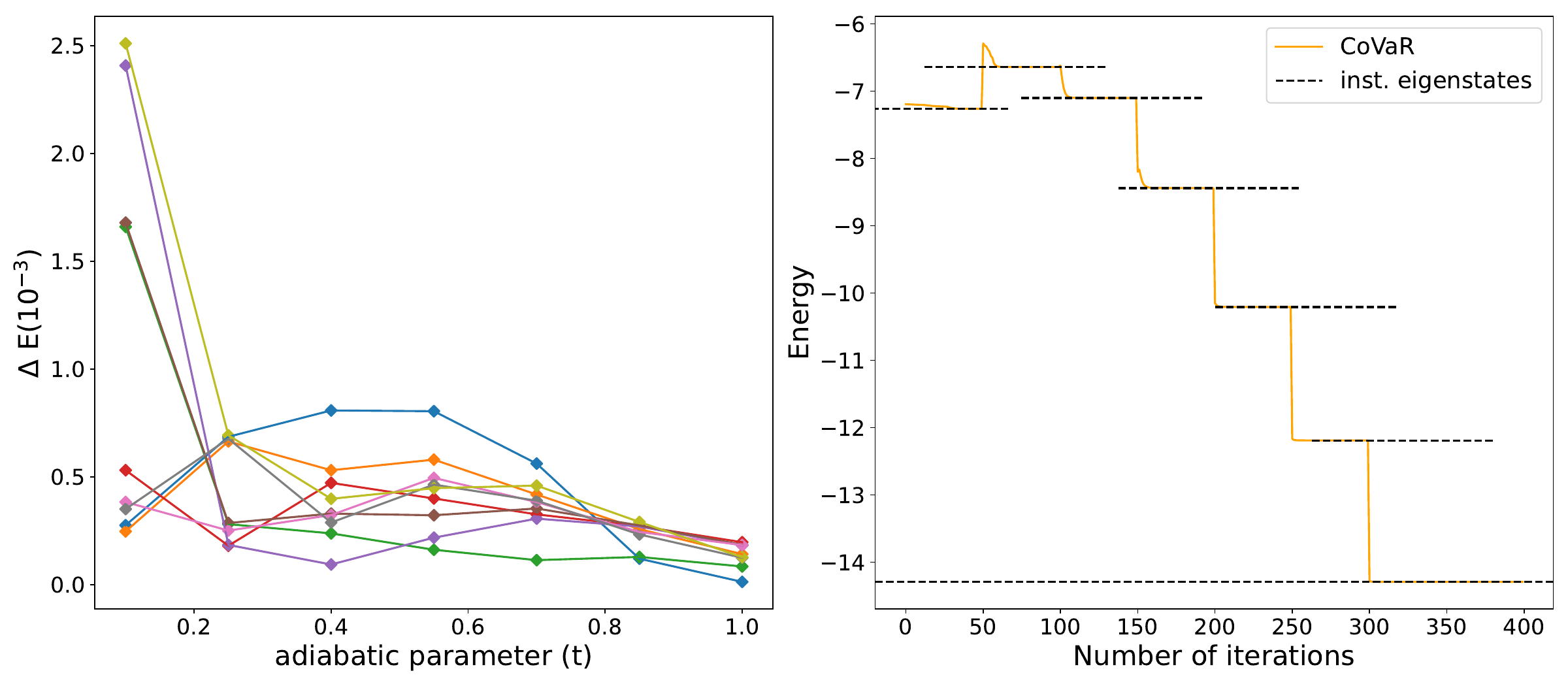} 
	\caption{
		Finding the ground state of an 8-qubit weighted max-cut problem Hamiltonian in \cref{eq:weight_mix}.
		(left) Energy difference from the actual ground state at each time-step ($\Delta t = 0.15$) when applying adiabatic CoVaR using a
		10-layer hardware-efficient ansatz.
		Each colour corresponds to a different randomly generated Hamiltonian whereby the number of distinct weights $w_{ij}$ was fixed to 14.
		A fixed number of 50 CoVaR iterations were performed at each time step.
		(right) Energy evolution (solid line) and instantaneous ground-state energies (dashed lines) of the time dependent Hamiltonian 
		in a single random, but representative instance of \cref{eq:weight_mix}. 
	}
	\label{fig:maxcut}
\end{figure*}

Second, we demonstrate how our approach extends ground-state preparations, as relevant in ref.~\cite{chakraborty2022classicallyemulateddigitalquantum}, to excited state
preparations by initialising our approach in low lying eigenstates of the initial model at $t = \Delta t$ (given the energy levels are degenerate at $t=0$). 
While, again, the CoVaR evolutions in \cref{fig:hep} (orange, purple, red) udergo energy excitations around small
intermediate energy gaps, the approach successfully prepares one of the eigenstates of the final model to high precision
(all final energies are within a distance $\Delta E \leq 10^{-3}$ from nearest energy eigenstates).

We also note that a clear difference from conventional adiabatic evolution can be observed in \cref{fig:hep} (left, red, $0.3 \leq t\leq0.4$),
whereby excitations to higher energy eigenstates happen in regions where the energy gaps are relatively large.
This is likely caused by closeness of those eigenstates in parameter space or high overlaps (fidelity) with the excited energy eigenstates.
Furthermore, CoVaR converged to the excited states with high precision in \cref{fig:hep} (right, purple, $0.6 \leq t\leq0.8$)
despite a large step size $\Delta t = 0.15$ and a number of nearby eigenstates with vanishingly small gaps -- in contrast phase estimation and conventional
adiabatic evolution would require 
long evolution times $T \propto g_{min}$ or $T \propto g_{min}^2$ and thus deep circuits when energy gaps are small.

\begin{figure*}[tb]
	\begin{center}
		\includegraphics[width=\textwidth]{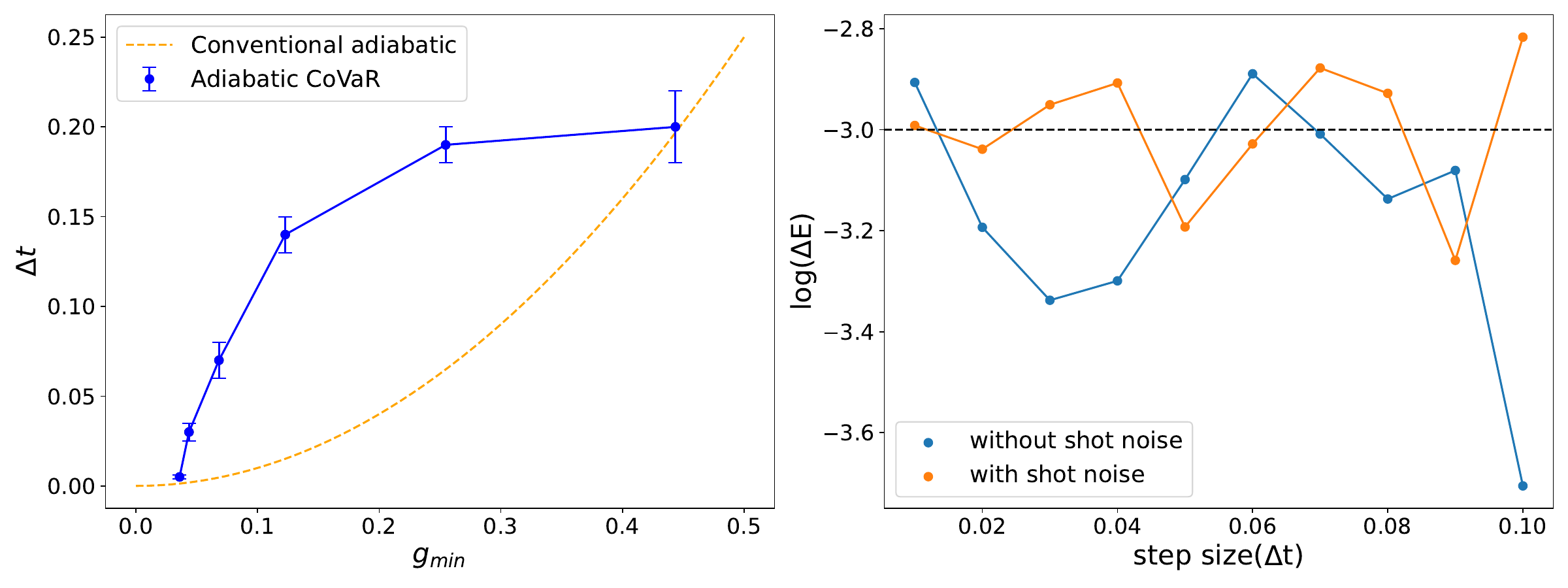} %
		\caption{
			(left) Largest time-step $\Delta t$ sufficient to achieve a final state preparation $\Delta E = \pm 0.0015$
			vs. minimum energy gap $g_{min}$ of the Hamiltonian.
			Error bars indicate variance across multiple randomly generated, 7-qubit spin Hamilonians from \cref{eq:spinring}. 
			For illustrative purposes, we plot (orange dotted line) the quadratic scaling of $\Delta t$ in conventional adiabatic evolution.
			This demonstrates that adiabatic CoVaR appears to have a fundamentally different dependence on the energy gap.
			(right) Final energy precision $\Delta E$ with (orange) and without (blue) shot noise
			confirming robustness of CoVaR against shot noise
			as long as the level of shot noise is slightly smaller than the sought final precision.
			The dashed line at $\Delta E = 10^{-3}$ indicates a target precision 
			that is sought in typical applications, e.g., chemical accuracy.
			\label{fig:noise_and_egap}
		}
	\end{center}
\end{figure*}

\subsection{Solving combinatorial optimisation problems}
We now demonstrate that adiabatic CoVaR can naturally be applied to approximating the 
solution to combinatorial optimisation
problems. We consider a weighted max-cut problem which can be encoded~\cite{farhi2014quantum,Lucas_2014}
as the ground state of the following Hamiltonian as
\begin{align}
	\label{eq:weighted_Hamil}
	\mathcal{H} = \sum_{i} w_i Z_i + \sum_{i<j} w_{ij} Z_i Z_j,
\end{align}
where we uniformly randomly generate the weights $0 \leq w_i \leq 1$ and $0 \leq w_{ij} \leq1$.
Given all terms in the Hamiltonian commute, and their common eigenstates are computational basis states,
one cannot apply the perturbtive approach. Instead, most commonly one uses a mixer Hamiltonian $\sum_{i} X_i$ 
to construct the time-dependent Hamiltonian as
\begin{align}
\label{eq:weight_mix}
	\mathcal{H}(t) =  (1{-}t)\sum_{i} X_i + t\{ \sum_{i} w_i Z_i + \sum_{i<j} w_{ij} Z_i Z_j\}.
\end{align}

\cref{fig:maxcut} (left) shows the difference from the instantaneous ground states 
for 10 different, randomly generated Hamiltonians, and demonstrates that CoVaR indeed closely follows instantaneous
ground states, with an energy difference $\Delta E \leq 2.5 \times 10^{-3}$.
It is worth noting that these simulations did not include shot noise in order to test the ultimate performance of the approach
and indeed we analyse the effect of shot noise separately in \cref{sec:shot_noise}.
We also note that the final precision achieved here ($\Delta E \leq 5 \times 10^{-4}$) is significantly higher than in
case of the spin problem in \cref{fig:CoVaR_vqe} given the ground state of the maxcut Hamiltonian
is a computational basis state which can be prepared exactly using our shallow ansatz.

To illustrate the effectiveness of CoVaR,
in \cref{fig:maxcut} (right) we plot the evolution of a single, representative random instance and
demonstrate that CoVaR (orange solid line) rapidly converges to the instantaneous ground states (black dashed lines).

\section{Further numerical analysis}

\subsection{Relationship between energy gap and step size}

In the conventional adiabatic time evolution approach a sufficiently
small timestep $\Delta t \sim g_{min}^2$ needs
to be chosen that is proportional to the square of the minimal energy gap
in order to prevent excitation to higher eigenstates~\cite{DJORDJEVIC2023491}.
In \cref{fig:noise_and_egap} (left)
we numerically estimate the largest feasible $\Delta t$ 
required to prepare the ground state of a series of randomly generated (by randomly sampling $c_i$) 7-qubit spin
problems [\cref{eq:spinring}] with different energy gaps $g_{min}$.

The value of $\Delta t$ was found through a linesearch whereby the largest $\Delta t$ was accepted that
achieved a final state preparation error $\epsilon \leq 0.0015$.
The error bars indicate variance across the different random Hamiltonians. 
Fitting a non-linear curve to the data in \cref{fig:noise_and_egap} (left)
suggests a logarithmic scaling $(\Delta t)^{-1} \sim \ln(g_{min})$ within the finite regime of $g_{min}$ explored.
This is in stark contrast to the conventional adiabatic evolution approach whose scaling is 
quadratic $\Delta t \propto g_{min}^2$, which we illustrate in \cref{fig:noise_and_egap} (left, dashed orange line).


\subsection{Robustness to shot noise \label{sec:shot_noise}}
Finally, we investigate the robustness of our approach to shot noise.
We use the approach of refs~\cite{Boyd_2022,boyd2024high} to simulate the effect of shot noise
by adding Gaussian noise of standard deviation $N_s^{-1/2}$ to individual covariances
and use a number of shots $ N_s= 10^6$.
We again consider the ground-state preparation of the 7-qubit spin Hamiltonian in \cref{eq:spinring}
using an ansatz of 16 layers. 

\cref{fig:noise_and_egap}(right) shows the final achieved energy error $\Delta E$ both with (orange) and without (blue) shot noise
for increasing values of $0.01  \leq  \Delta t \leq 0.1$.
Comparing the blue and orange curves it is clear that the present level of shot noise
does not significantly influence the performance of the approach.

These results nicely confirm observations of ref.~\cite{Boyd_2022} whereby a similar
robustness of CoVaR against shot noise was observed
and whereby the primary limitation due to shot noise was fund that
the final precision reached by CoVaR gets limited to a level determined by shot noise as $\Delta E \propto N_s^{-1/2}$,
and here $\Delta E \approx 10^{-3}$.

\section{Discussion and Conclusion}

	We consider CoVaR~\cite{Boyd_2022} which allows one to find eigenstates of a Hamiltonian;
	Parameters of an ansatz circuit are guided through a root-finding approach to ultimately 
	find joint roots of a large number of covariance functions.
	Joint roots guarantee that the quantum state prepared by the parametrised circuit is an eigenstate of the
	problem Hamiltonian. A large number of covariances $\propto 10^8$--$10^{10}$ can be extracted efficiently
	for large systems using classical shadows, guaranteeing remarkable robustness and convergence speed of the approach. 
	However, CoVaR is oblivious as to which eigenstate it converges to and thus its vanilla 
	implementation required an initial state that has high overlap with the desired target eigenstate -- an assumption
	most typical in quantum simulation.

	In the present work we attempt to address this limitation and combine the powerful CoVaR approach with a discretised morphing
	of an initial trivial Hamiltonian into the final desired problem Hamiltonian. By design, the initial trivial Hamiltonian's eigenstates
	can be solved analytically; the approach is then efficiently initialised in one of these eigenstates and, after slightly adjusting the 
	Hamiltonian as in adiabatic computing, CoVaR is employed to prepare the instantaneous eigenstates of the time-dependent Hamiltonian. This
	way the initial state supplied to the CoVaR subroutine has a high overlap with the target state.

	We numerically simulated the present approach in a broad range of practically relevant applications.
	First, we considered a perturbative approach whereby an initial trivial Hamiltonian is morphed by gradually turning on a strong coupling term.
	We compared adiabatic CoVaR in a 10-qubit spin chain system to two different implementations of VQE.
	a) by performing standard VQE using the same initial parameters that prepare the ground state
	of the initial trivial model.
	b) we performed an adiabatic VQE by effectively using VQE instead of CoVaR in the present morphing approach.
	All instances of these VQE optimisations failed and got stuck in local minima whereas the present approach
	converged to the true ground state in all instances with relatively high precision.
	
	Second, a broad application area where adiabatic techniques are very naturally applicable are
	combinatorial optimisation problems~\cite{farhi2014quantumapproximateoptimizationalgorithm}.
	We apply adiabatic CoVaR to solving weighted max-cut problems 
	demonstrate its effectiveness by successfully preparing
	exact ground state in all instances.
	
	A significant strength of adiabatic CoVaR is that it naturally allows for the direct preparation
	of excited states similarly to the phase estimation algorithm -- in contrast in VQE variants, such as VQD, one needs to
	prepare excited states one-by-one.  We demonstrated excited-state preparation in two practically relevant application examples:
	in preparing highly excited states in a random Heisenberg model and in the Schwinger model which is a toy model relevant in
	High-Energy-Physics applications. Although our approach is demonstrably robust against local traps and did in all
	simulations converge to an eigenstate, depending on the choice of $\Delta t$, small energy gaps
	typically led to convergence to one of the nearby eigenstates in parameter space, effectively resulting in an
	excitation or de-excitation of the variational state.

	However, the most remarkable feature of adiabatic CoVaR is that it appears to tolerate small
	or vanishing energy gaps in the following sense. We numerically demonstrated that, while
	one does not have full control as to which eigenstate CoVaR converges to, the present approach successfully
	prepares an eigenstate even when the energy gaps are quite small
	and the time step is very large as $\Delta t = 0.15$.
	In one example we empirically found that the adiabatic time step $\Delta t$
	scaled logarithmically with the minimum energy gap $g_{min}$ of the problem Hamiltonian.	
	This is a significant and unique advantage of the present approach as even 
	phase estimation---applied to project to instantaneous eigenstates of the time-dependent Hamiltonian---requires
	deep quantum circuits due to long evolution times $T \propto g_{min}^{-1}$
	when energy gaps are too small.
	Similarly, this dependence is quadratic as $T \propto g_{min}^{-2}$
	in conventional adiabatic evolution.

	Finally, we analysed the robustness of the present approach against imperfections and against variability in
	hyperparameter settings.
	Our numerical results confirm prior observations that CoVaR is robust against shot noise.
	Furthermore, ref.~\cite{Boyd_2022} confirmed that CoVaR is robust against gate noise 
	given the perturbations to the ideal Jacobian matrix due to gate imperfections are averaged out due to the
	large dimensionality of the Jacobian matrix
	-- and also because the update rule in~\eqref{eq:parameter_update} is invariant under
	averaged out global depolarizing noise. Still, we note that a
	broad range of error mitigation techniques~\cite{Cai_2023, PhysRevX.11.031057, Koczor_2021} are immediately applicable
	to CoVaR through error mitigated classical shadows~\cite{PRXQuantum.5.010324}.
		
	 Finally, we note that we focused on conventional ``Pauli shadows''~\cite{Huang_2020} which are straightforward to implement. 
	 However, a broad range of more advanced shadow measurement techniques are available that are immediately applicable
	 to extend the range of the present approach or to make it more shot efficient 
	 including shallow shadows~\cite{bertoniShallowShadowsExpectation2022} or fermionic shadows~\cite{wanMatchgateShadowsFermionic2023, wuErrormitigatedFermionicClassical2024}
	 -- while the latter enables to study ground and excited states of molecular and materials systems~\cite{McArdle_2020, oled, PhysRevResearch.4.013052, WANG20242877}.


\section*{Data Availability}
The core algorithm presented in this work is fully implemented and demonstrated in the repository
openly available at \url{https://github.com/seop02/adiabatic_covar/tree/main}.

\section*{Acknowledgments}
The authors thank Bipasha Chakraborty for helpful discussions.
BK thanks the University of Oxford for a Glasstone Research Fellowship and Lady Margaret Hall, Oxford for a Research Fellowship.
All simulations were performed using the open-source tools Quantum Exact Simulation Toolkit
(QuEST)~\cite{Jones2019} and its Mathematica-based interface QuEST link~\cite{Jones_2020}.
The authors also acknowledge funding from the
EPSRC projects Robust and Reliable Quantum Computing (RoaRQ, EP/W032635/1)
and Software Enabling Early Quantum Advantage (SEEQA, EP/Y004655/1).

\appendix
\section{Comparison to prior techniques}

\begin{figure}[tb]
	\centering
	\includegraphics[width=0.45\textwidth]{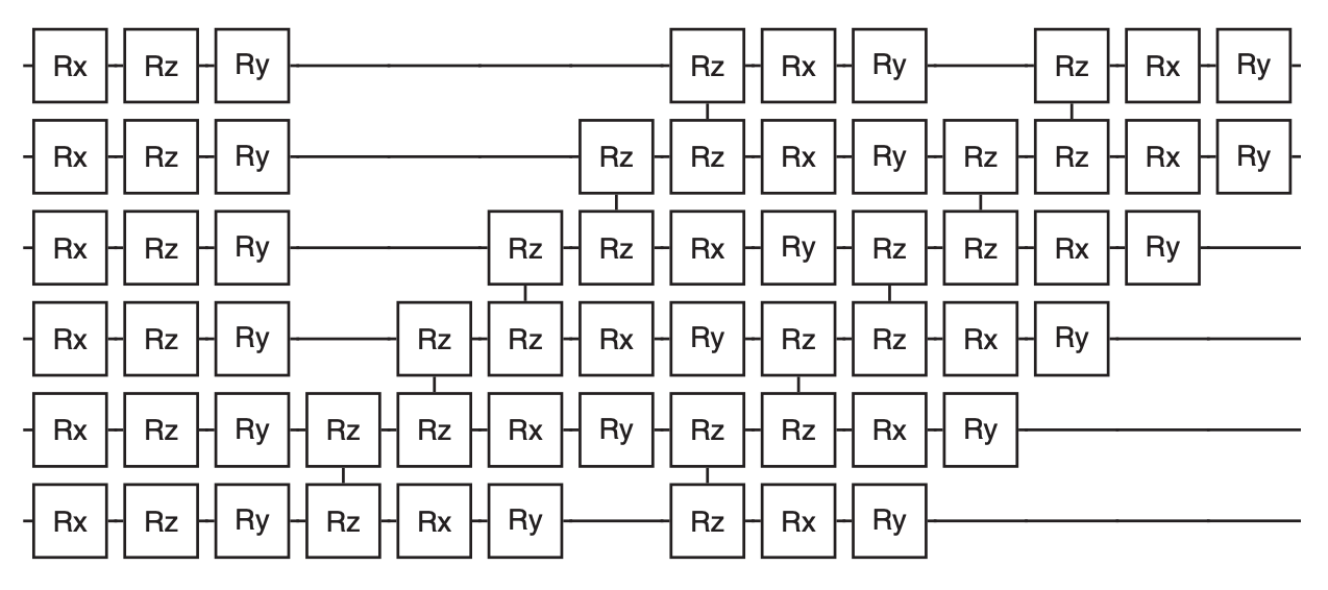}
	\caption{Example of a single layer of a 6-qubit hardware-efficient ansatz used in the present work.}
	\label{fig:ansatz}
\end{figure}

\begin{figure*}[t]
	\centering\includegraphics[width=\textwidth]{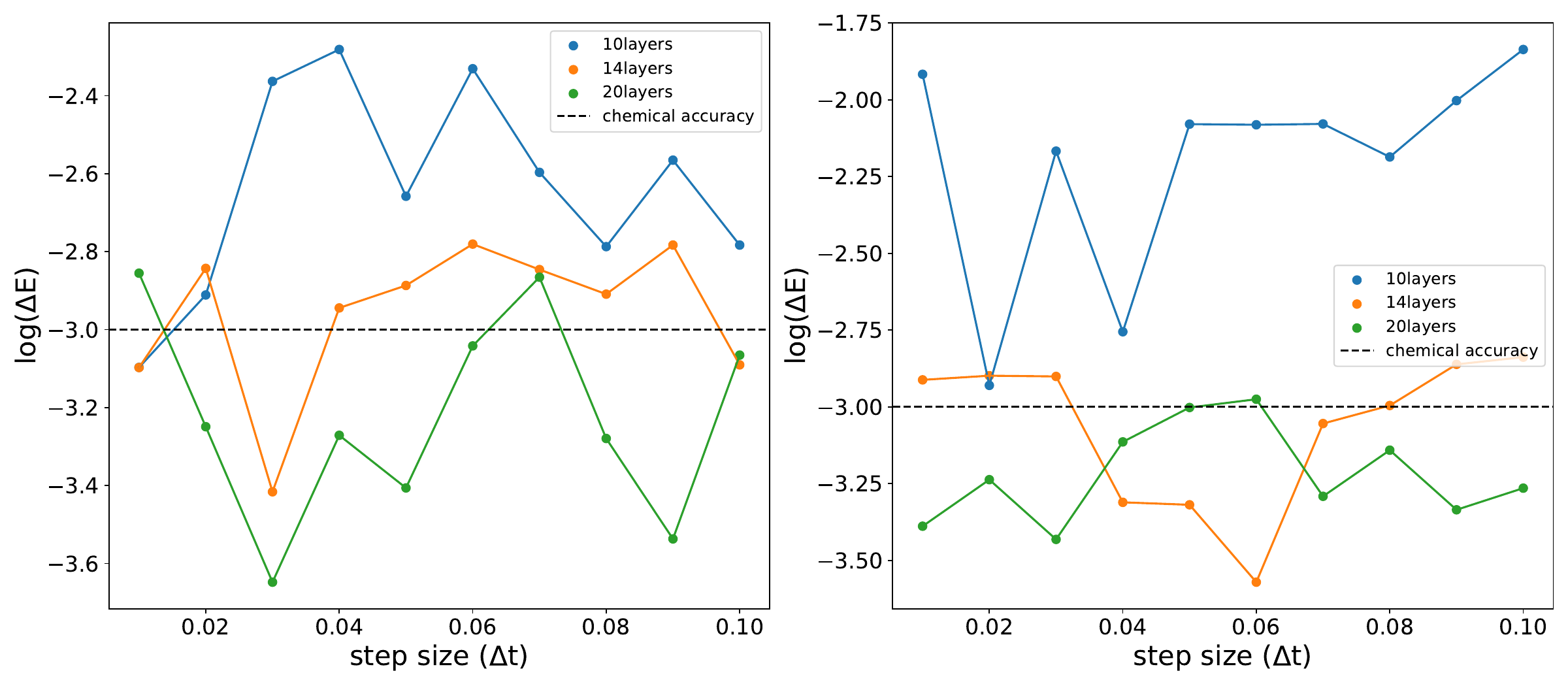}
	\label{fig:diff_all}
	\caption{
		Final energy difference from the exact ground (left) and first excited state (right)
		for an increasing number of ansatz layers and an increasing
		step size $\Delta t$ in a 7-qubit spin Hamiltonian from \cref{eq:spinring}.
		The black horizontal dashed line represents a typical target precision, e.g., chemical accuracy.
	}
	\label{fig:precision}
\end{figure*}

\subsection{Variational Quantum Deflation \label{app:vqd}}
Variational Quantum Deflation (VQD) is a variant of VQE
for preparing excited states~\cite{Higgott_2019, Jones_2019}.
The approach augments the VQE cost function with additional terms
that account for the overlap with known energy eigenstates, i.e.,
the VQD cost function for finding the $k$-th excited state is
\begin{align*}
	C_k (\boldsymbol{\theta}_k) = \bra{\psi(\boldsymbol{\theta}_k)} \mathcal{H} \ket{\psi(\boldsymbol{\theta}_k)} + \sum_{i = 0} ^ {k-1} \beta_{i} |\bra{\psi(\boldsymbol{\theta}_k)} \ket{\psi(\boldsymbol{\theta}_i)}|^2.
\end{align*}
Here $\beta_{i}$ is an arbitrary hyperparameter and
$\boldsymbol{\theta}_i$ are parameter settings that
approximate the $i$-th excited energy eigenstate $\ket{\psi_i}$.

This ensures that we minimize the energy with the constraint
that the variational state be orthogonal to
the previously found eigenstates~\cite{vqe_excited}.
The downside of the approach is that one requires explicit
knowledge of the parameter settings
$\boldsymbol{\theta}_0, \dots , \boldsymbol{\theta}_{k-1}$
up to the $(k-1)^{th}$ excited state in order to find the $k$-th excited state, which
in practice is achieved by iteratively finding higher and higher excited states.

For this reason, preparing highly excited states
may be challenging with VQD given the states are prepared incrementally 
and thus errors in the lower excited states are
expected to accumulate, i.e., at each step one is vulnerable to barren plateaus 
and to getting trapped in local minima.


\section{Analysing dependence on hyper parameters}

\subsection{Effect of adiabatic time-step and circuit layers on the precision}
\label{sec:step&layer}

In all simulations in the present work we used a hardware efficient ansatz illustrated in \cref{fig:ansatz}.
Here, we present how the number of circuit layers influences the precision of the state preparation.
Simulation results for both ground state energy and first excited states are presented for
a 7-qubit spin problem from \cref{eq:spinring} in \ref{fig:precision} for an increasing number of layers
between 10 and 20 and for an increasing step size $\Delta t$. 

While 10 variational layers in \cref{fig:precision} (left and right) are not sufficient for achieving a target precision
of $10^{-3}$, increasing the number of layers to $14$ and $20$ clearly improves the performance.
Furthermore, it appears the performance of adiabatic CoVaR is fairly robust against the step size $\Delta$ as increasing
$\Delta$ does not significantly affect the final energy difference.

Increasing the number of parameters does indeed improve the precision, however, comes with the trade-off of an increased circuit depth
and also an increased runtime through an increased number of circuit runs. Thus, in practice it is worth
minimising the number of layers as a hyper parameter.

Here, we present how the number of parameters in the variational circuit, specifically the number of variational layers, influences the precision of the final energy. Simulation results for both ground state energy and first excited states are presented in Figure \ref{fig:precision}. All simulations in this section were conducted on a 7-qubit system, with the number of layers ranging from 10 to 20. We illustrate results for 10, 14, and 20 layers, which were sufficient to illustrate the trend.

The figures ~\ref{fig:precision} immediately reveal that 10 variational layers are insufficient for a 7-qubit system. For the first excited state computation, it does not achieve chemical accuracy for all adiabatic runtimes($\Delta t$s). For the ground state computation, chemical accuracy is only achieved for $\Delta t = 0.01$. Precision notably improves for 14 and 20 layers. In simulations with 14 layers, chemical accuracy is attained for the ground state when $\Delta t = 0.03$ and $\Delta t = 0.1$. For the first excited state, chemical accuracy is achieved for $\Delta t$ ranging from $\Delta t = 0.03$ to $\Delta t = 0.08$. More accurate results are obtained with 20 layers. For the ground state computation, chemical accuracy is achieved for almost all $\Delta t$ values except for $\Delta t = 0.01$ and $\Delta t = 0.07$. For the first excited state computation, chemical accuracy is achieved for almost all $\Delta t$ values except for $\Delta t = 0.06$.

The results imply that the precision of the final energy is heavily dependent on the number of parameters within the variational circuit. Increasing the number of parameters does tend to improve the precision. However, we have to note that there exists a trade-off between the number of parameters and the computation run-time. More parameters result in increased computational time for each CoVaR iteration, as we would have to compute more covariances. To optimize the computation process, it is important to find the most optimum number of layers that reach the chemical accuracy while minimizing the computation time.


\subsection{Effect of adiabatic time-step and circuit layers on the number of total iterations}
\label{sec:step&iter}

\begin{figure*}[t]
    \centering\includegraphics[width=\textwidth]{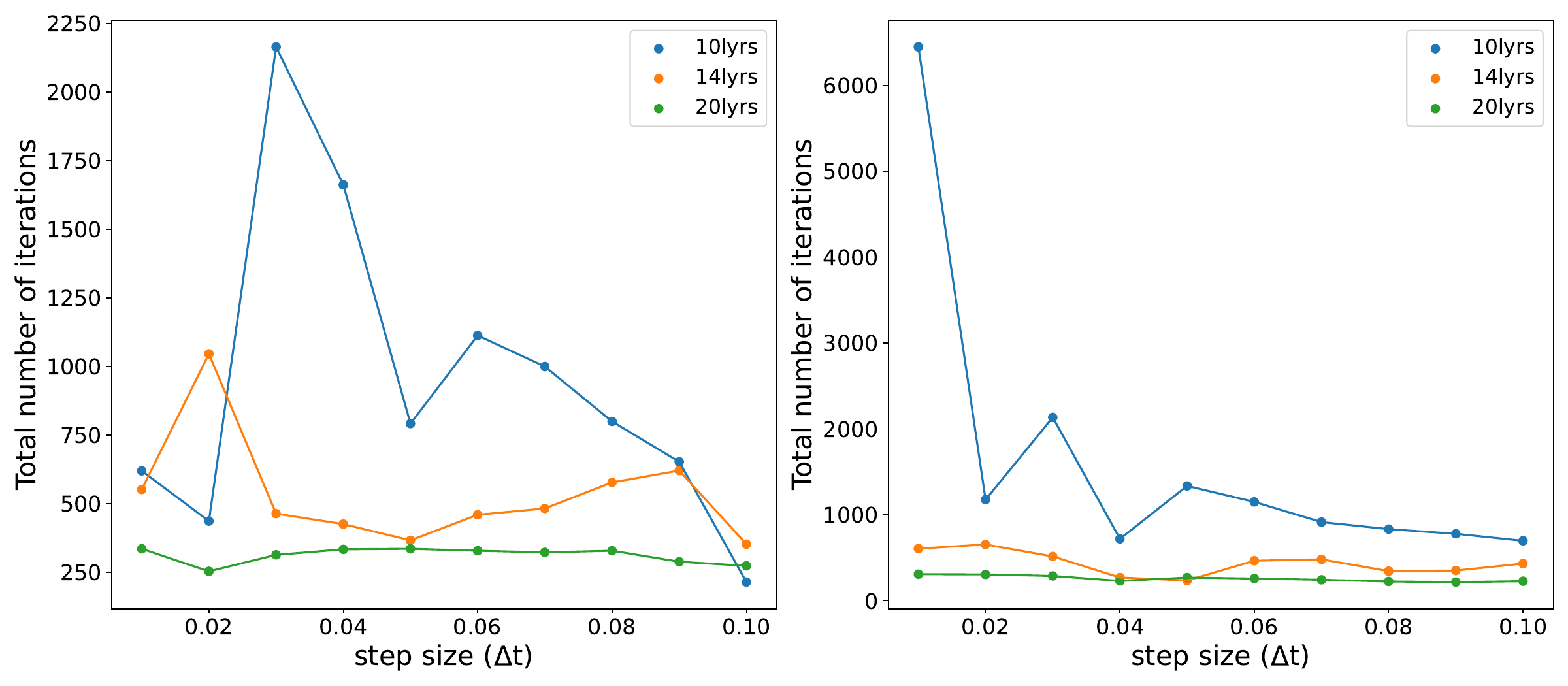}
     \label{fig:diff_counts}%
     
     \caption{Total number of CoVaR iterations required to reach the final ground state (left) and the first excited state (right). 
     	Here we present the total number of CoVaR iterations throughout the entire adiabatic procedure.
     	}
     \label{fig:counts}
\end{figure*}

The total number of CoVaR iterations used for the entire adiabatic morphing is strongly influenced by the expressivity of the ansatz.
In particular, in the present implementation of adiabatic CoVaR we demand that for each adiabatic timestep
CoVaR must converge to the instantaneous eigenstate to precision $\epsilon = 0.002$ in the covariance vector norm.
Passing this threshold, CoVaR terminates, and the computation proceeds to the next time step.

\begin{figure*}[tb]
	\centering\includegraphics[width=\textwidth]{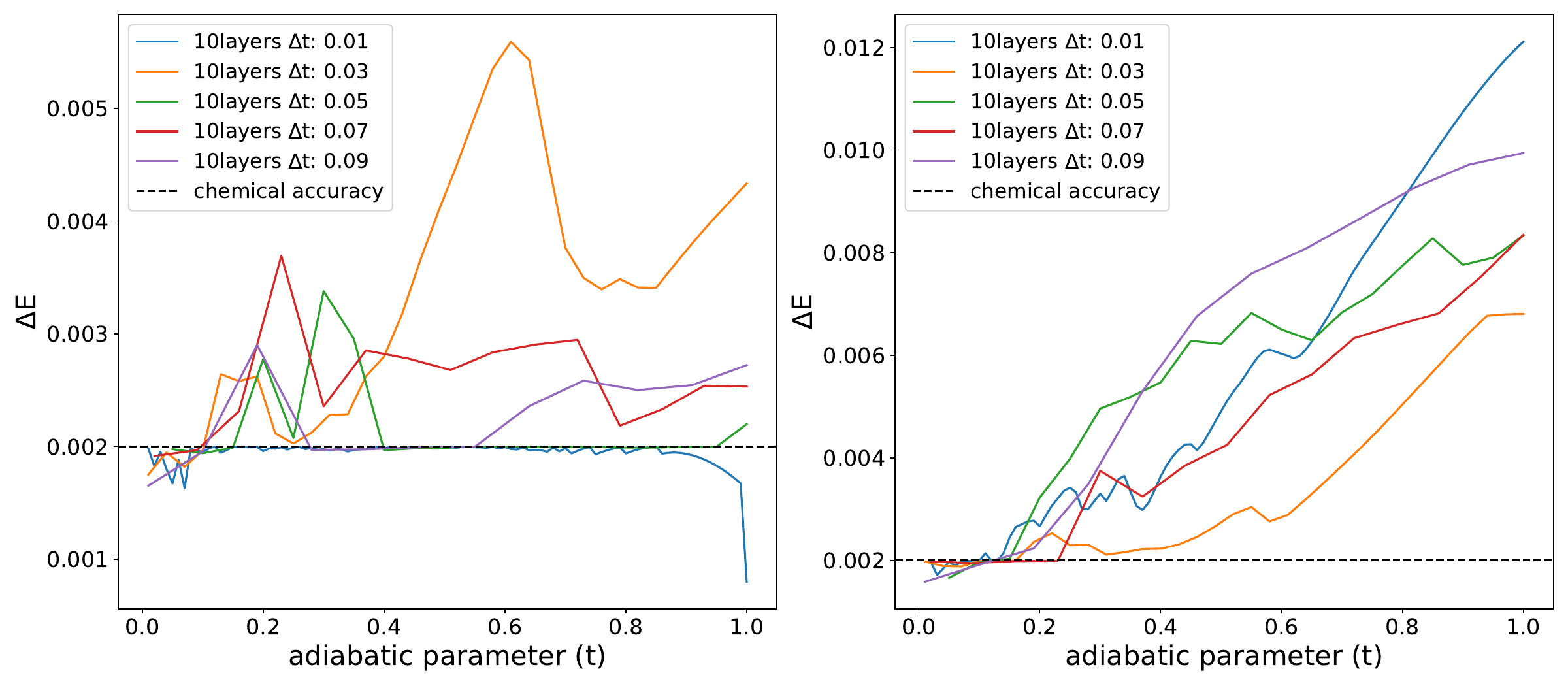}
	\label{fig:diff_evol}%
	\caption{Evolution of the energy difference with 10 variational layers for the ground (left) and the first excited states (right)
		throughout the adiabatic procedure $t \in [0,1]$, with a threshold line indicating the termination criterion for CoVaR.
		While the ground state preparation (left) demonstrates consistent convergence, the first excited state (right) 
		is prepared to a decreasing precision due to the insufficient circuit depth in the variational ansatz.
	}
	\label{fig:energy_diff_exc_state}
\end{figure*}

\cref{fig:counts} illustrates that circuits with 10 layers required the highest number of CoVaR iterations
to reach the final eigenstate in contrast to circuits with more layers. The reason is likely that the shallow
variational state was not able to sufficiently well express the instantaneous eigenstates throughout the evolution and 
thereby the threshold condition was not always passed as further elucidated in \cref{fig:energy_diff_exc_state}. As the number of layers is increased, the average number of CoVaR
iterations decreases significantly.

Our analysis reveals a complex interplay between run-time, iteration count, and layer number in quantum computations. As run-time decreases due to increasing $\Delta t$, the total number of iterations generally decreases across all layers, an expected outcome given the reduced maximum possible iterations with larger $\Delta t$ values. However, we observe a notable shift in this trend when $\Delta t$ exceeds 0.05, where iterations begin to increase for both ground state and first excited state calculations, suggesting an optimal $\Delta t$ for iteration minimization. In the case of first excited state computations, the 10-layer configuration required significantly more iterations compared to other layer counts, while 14 and 20 layers exhibited similar iteration counts. This similarity suggests that 14 layers may be sufficient to achieve energy within chemical accuracy, although it's important to note that similar iteration counts do not necessarily imply equivalent computation times due to the increased parameter count in the 20-layer configuration. The increase in iterations at larger $\Delta t$ values can be attributed to the initialization of each CoVaR computation farther from the target eigenstate, necessitating more iterations for convergence. Intriguingly, we found that the optimal $\Delta t$ for minimizing iterations coincides with the precision required to achieve chemical accuracy, a finding that merits further investigation in future studies.


\section{Details of \cref{fig:CoVaR_vqe}}
\label{sec:CoVaR_vqe}

A 30-layer hardware-efficient ansatz was used with $\Delta t = 0.05$.
Performing
a single iteration of CoVaR is nearly of the same quantum cost as performing an iteration of VQE.
For a fair comparison, 50 iterations were performed for $t<1$,
and 100 iterations at $t=1$ for the adiabatic CoVaR (blue) and adiabatic VQE (orange)
while the same number $1050$ of total iterations were used for conventional VQE. 

\begin{figure}[tb]
    \centering
    \includegraphics[width=0.47\textwidth]{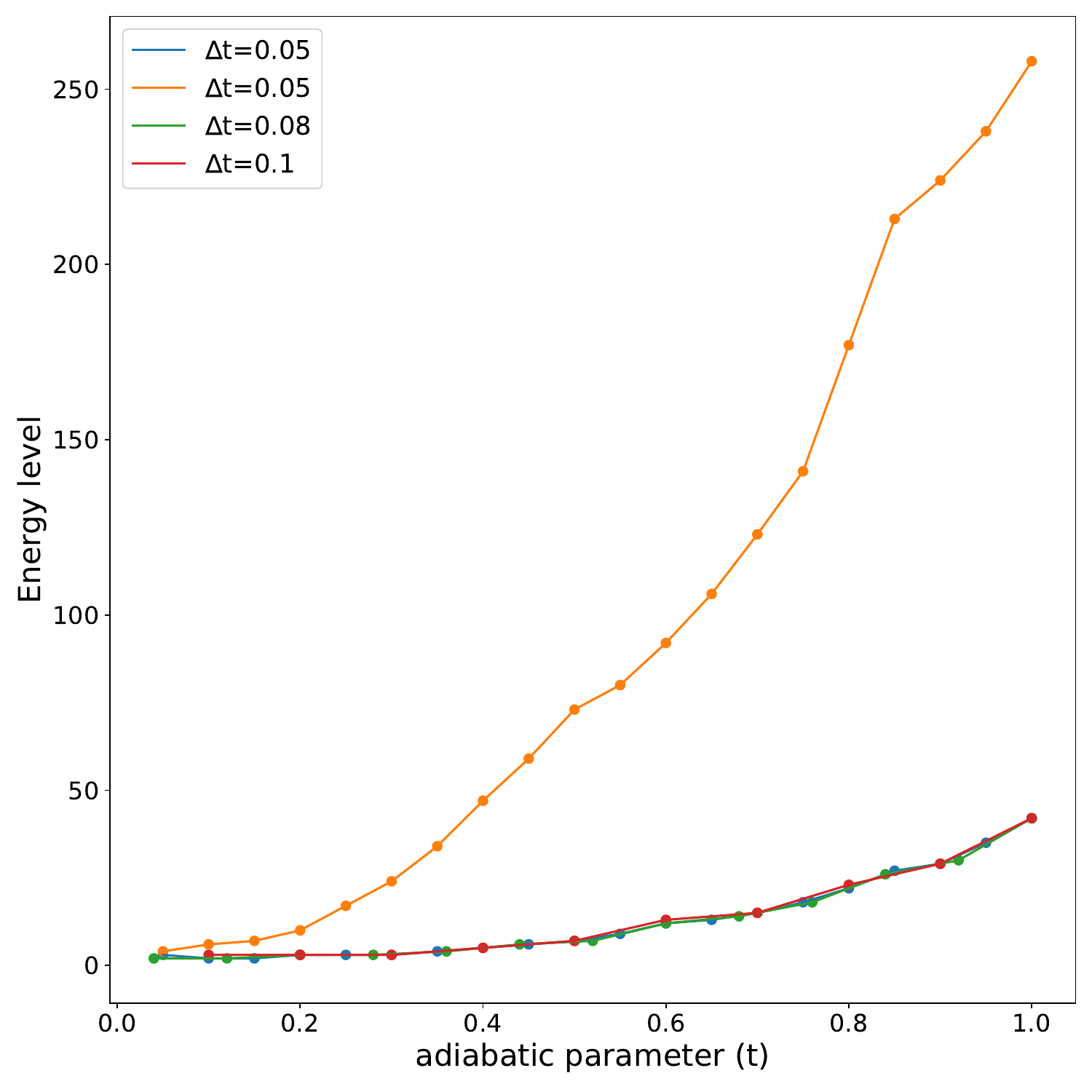}
\caption{
	Index of the energy levels obtained during adiabatic CoVaR in \cref{fig:CoVaR_vqe} (right).
	}
    \label{fig:level}
\end{figure}

\cref{fig:level} presents the index of energy levels obtained  initialising adiabatic CoVaR in the first excited state 
of the initial Hamiltonian in \cref{fig:CoVaR_vqe} (right).
We performed to different simulations using the time step $\Delta t = 0.05$,
and in one case the energy levels were more rapidly increased than in case of all other simulations.
This can be primarily attributed to the random nature of the initial variational parameter in CoVaR.
The randomness inherent in CoVaR maz lead to significant differences in the evolutionary trajectories. 
Demonstrably, even with identical hyper parameters, the resulting trajectories are radically different.

\section{Details of \cref{fig:energy_diff_exc_state}}
\label{sec:ediff_evol}

Here, we present the evolution of the energy difference between the variational state and the target eigenstate with insufficient number of variational layers. A 10-layer hardware-efficient ansatz was used for computations of both ground state and the first excited state in the spin chain model. In order to illustrate the effect of insufficient variational layers on the precision of the final energy obtained by adiabatic CoVaR, we fixed the number of variational layers and ran multiple simulations varying the adiabatic runtime($\Delta t$).

\cref{fig:diff_evol} depicts the energy difference evolution for the spin chain system. For the ground state computation, most runtimes failed to achieve chemical accuracy, with $\Delta t = 0.01$ being the sole exception. The first excited state computations fared worse, with all runtimes failing to reach chemical accuracy and the energy difference diverging throughout the adiabatic evolution. These results stem from our use of a coupling constant J = 1.0, which led to a highly entangled and mixed final state, markedly different from the initial unperturbed eigenstate. This increased entanglement presented a formidable challenge for eigenstate computation, necessitating greater circuit depths and resulting in a gradual increase in energy deviation. Furthermore, the adiabatic evolution of the Hamiltonian caused a shift in the system's eigenstates, expanding the Hilbert space that the variational ansatz needed to span. This expansion further underscored the need for larger variational depths in adiabatic CoVaR. These findings highlight the crucial importance of carefully selecting both the number of variational layers and runtime parameters in adiabatic CoVaR computations, especially when dealing with highly entangled systems. To address these challenges, future research could explore adaptive layer techniques or alternative ansatz structures, potentially offering more robust solutions for these complex scenarios.

\bibliography{ref}

\end{document}